\documentclass[11pt,fleqn]{article}
\usepackage{graphicx}
\usepackage[a4paper,left=2cm,right=2cm]{geometry}
\usepackage{amsmath}
\usepackage{float}
\usepackage[latin1]{inputenc}
\usepackage{amsmath}
\usepackage{amsfonts}
\usepackage{amssymb}
\usepackage{float}
\floatstyle{boxed} 
\restylefloat{figure}

\newcommand{\be}{\begin{equation}}
\newcommand{\ee}{\end{equation}}
\newcommand{\bea}{\begin{eqnarray}}
\newcommand{\eea}{\end{eqnarray}}

\begin{document}
\title{A simple extension of SM that can explain the $(g-2)_{\mu}$ anomaly, small neutrino mass and a dark-matter.}
\author{Lobsang Dhargyal. \\\ Institute of Mathematical Sciences, HBNI, Chennai 600113 India \\\ and \\\ Harish-Chandra Research Institute, HBNI, Chhatnag Road, Jhusi Allahabad 211 019 India.}

\maketitle
\begin{abstract}

In this work we propose a simple extension of standard-model (SM) by adding eleven new particles to it. Three heavy leptons ($f_{e},\ f_{\mu},\ f_{\tau}$) singlet under the $SU(3)_{c}\times SU(2)_{L}$ carrying respective Lepton-Numbers, charged under the $U(1)_{Y}$ with $Y = -2$ and transforming under a discrete symmetry as $f_{i} \rightarrow -f_{i}$. One scalar ($\phi_{2}$), singlet under all the SM gauge groups and transforming under the discrete symmetry as $\phi_{2} \rightarrow -\phi_{2}$ which does not develops a non zero vacuum-expectation-value (VEV). One more scalar ($\phi_{3}$), singlet under all the SM gauge groups and invariant under the discrete symmetry which develops a non zero VEV ($v_{3}$) and gives masses to $f_{i}$s, $\phi_{2}$ and neutrinos. Three right-handed neutrinos ($\nu_{iR}$) and three left-handed Majorana neutrinos ($s_{iL}$). With these new additional particles added to SM we have been able to give explanations to the long standing muon (g -2) anomaly as well as the smallness of neutrino masses by the inverse see-saw mechanism. And also in this model we have a very suitable scalar dark-matter (DM) candidate in $\phi_{2}$ with allowed mass as high as 53 GeV, although due to large Yukawa coupling required to explain the muon (g-2), its contribution to the DM relic density turn out to be too small and so it can account only a small fraction of the DM relic density of the universe.

\end{abstract}

Minor typos corrected to "J.Phys. G45 (2018) no.7, 075002. DOI: 10.1088/1361-6471/aac40a" in Table 1.

\section{\large Introduction.}

Dirac equations for a charged spin half muon predicts a magnetic moment, $\vec{M} = g_{\mu}\frac{e}{2m_{\mu}}\vec{S}$, with the gyromagnetic ratio $g_{\mu} = 2$. But quantum loop effects leads to small calculable correction to the $g_{\mu}$, which make its true value deviates from 2, parametrized by the anomalous magnetic moment of muon
\be
a_{\mu} = \frac{g_{\mu} - 2}{2}.
\ee
The $a_{\mu}$ can be accurately measured and in a given model such as Standard Model (SM), it can be precisely predicted. Hence $a_{\mu}$ is a very good laboratory to test SM predictions at its quantum loop level. Any deviation from SM prediction would signal presence of New Physics (NP), with current sensitivity reaching up to mass scale of $\mathcal{O}$(TeV) \cite{pdg1}\cite{pdg2}. The latest PDG world average of $a_{\mu}^{Exp}$ is \cite{pdg-muon}
\be
a_{\mu}^{exp} = 11659209.1(5.4)(3.3)\times 10^{-10},
\ee
and combining the QED, EW and Hadronic parts together we have the SM prediction of $a_{\mu}$ \cite{pdg-muon}
\be
a_{\mu}^{SM} = 116591803(1)(42)(26)\times 10^{-11},
\ee
where the errors are due to the EW, lowest-order hadronic, and higher-order hadronic contributions, respectively. The difference between the experimental value and the SM value is
\be
\delta a_{\mu} = a_{\mu}^{exp} - a_{\mu}^{SM} = 288(63)(49)\times 10^{-11},
\label{diff-a}
\ee
and when all the errors are added in quadrature, the deviation of the experimental value from the SM prediction amount to 3.6 $\sigma$ \cite{pdg-muon}. Beside the above anomaly in muon (g-2), there is the long standing problem in SM of the unnatural smallness of neutrino masses and the existence of DM. One of the simplest way to understand the smallness of neutrino masses is through the inverse-seesaw mechanisms, where existence of heavy Majorana neutrinos and lepton number conservation can suppress the masses of the SM light neutrinos. The simplest of inverse-seesaw can be realized with three SM singlet right handed neutrino partners ($\nu_{Ri}$) to the three left handed neutrinos of SM along with three left handed Majorana neutrinos ($s_{Li}$), where index $i$ refer to the three generations of leptons. There has been many NP models proposed to explain the anomaly in $a_{\mu}$, one of them being the contribution to it from supersymmetry particles \cite{pdg1}, another NP model proposed is the ``dark photon" scenarios given in \cite{pdg33}-\cite{pdg35}. It has also been shown in \cite{Abe7}\cite{Abe8} that the lepton specific (Type-X) two-Higgs-doublet model (2HDM) can give significant enhancement of $a_{\mu}$. More recently in \cite{Abe}, it has been shown that in a muon specific 2HDM, the anomaly can be reduced within 1$\sigma$ provided $\tan{\beta}$ is very large. For more comprehensive coverage see \cite{Jheler} and the recent review by the PDG group \cite{pdg-muon} and the references there in. In this work we introduce a new SM singlet scalar $\phi_{2}$ stabilized by $Z_{2}$ symmetry which can be a DM candidate, but due to requirement of large Yukawa coupling to explain the muon (g-2), its contributions to the present relic density as estimated in \cite{Abe}\cite{Chaing} in models similar to ours (in DM annihilation calculation), turn out to be too small for it to constitute the whole DM mass of the universe.

\section{Model.}

\subsection{\large Lagrangian.}

To SM we add eleven new particles, three $SU(3)_{c}\times SU(2)_{L}$ singlet heavy lepton ($f_{e}, f_{\mu}, f_{\tau}$) transforming under a discrete symmetry as $f_{i} \rightarrow -f_{i}$ and charged under the SM $U(1)_{Y}$ with Y = -2 carrying respective Lepton-numbers. One $SU(3)_{c}\times SU(2)_{L}\times U(1)_{Y}$ singlet scalar ($\phi_{2}$) also transforming under the discrete symmetry as $\phi_{2} \rightarrow -\phi_{2}$, whose Vacuum-Expectation-Value (VEV) is zero. And one more $SU(3)_{c}\times SU(2)_{L}\times U(1)_{Y}$ singlet scalar ($\phi_{3}$) transforming under the discrete symmetry as $\phi_{3} \rightarrow +\phi_{3}$, which develops a non zero VEV $v_{3}$, and gives masses to $f_{i}$ and $\phi_{2}$ and neutrinos. To explain the unnatural smallness of neutrino masses, we introduce three SM singlet right handed partners ($\nu_{iR}$) to the SM left handed neutrinos and three SM singlet left handed Majorana neutrinos ($s_{iL}$), where index $i$ refer to three generations. All SM particles are invariant under the discrete symmetry transformation. The new particles and their transformation properties are given in the Table \ref{tab1}.

\begin{table}[h!]
\begin{center}
\begin{tabular}[b]{|c|c|c|c|c|} \hline
Particles & $SU(3)_{c}$ & $SU(2)_{L}$ & $U(1)_{Y}$ & $Z_{2}$ \\
\hline\hline
$\phi_{2}$ & 1 & 1 & 0 & -1 \\
\hline
$\phi_{3}$ & 1 & 1 & 0 & +1 \\
\hline
$f_{iL}$ & 1 & 1 & -2 & -1 \\
\hline
$f_{iR}$ & 1 & 1 & -2 & -1 \\
\hline
$\nu_{iR}$ & 1 & 1 & 0 & -1 \\
\hline
$s_{iL}$ & 1 & 1 & 0 & -1 \\
\hline
\end{tabular}
\end{center}
\caption{The new particles and their transformation properties under the SM gauge groups and $Z_{2}$.}
\label{tab1}
\end{table}

Then the Yukawa sector which is invariant under all the above symmetry transformations is given as
\be
\mathcal{L}_{Yukawa} = \sum^{\tau}_{i = e}Y_{2i}\bar{\mu}_{Ri}f_{Li}\phi_{2} + \sum^{\tau}_{i = e}Y_{3i}\bar{f}_{Ri}f_{Li}\phi_{3} + h.c.
\label{Yukawa}
\ee
Now constrains from heavy charged lepton searches \cite{heavy-charge-search}, which is mainly focus on $f^{\pm} \rightarrow \nu W^{\pm}$ or $f^{\pm} \rightarrow l^{\pm} Z$ decay types, has ruled out $m_{f} \le 100.8$ GeV at 95\% CL. So from Eqs.(\ref{Yukawa}), it is clear that if $\phi_{2}$ develops VEV, then f will mix with $\mu$ to diagonalize the mixed mass generated from the first term in the Eqs.(\ref{Yukawa}) and therefore the bonds from heavy charged lepton searches applies. But if $\phi_{2}$ does not develops VEV, then heavier lepton f will not mix with the $\mu$ and the bounds from the search for heavy charged leptons does not apply as it can not decay into the final states that those experiments looked for.\\
Then the general Lagrangian invariant under all the symmetry transformations for the new particles can be written as
\be
\mathcal{L}_{NP} = \mathcal{L}_{Kinetic} - V(H,\phi_{2},\phi_{3}) + \mathcal{L}_{Yukawa}
\ee
where
\be
\mathcal{L}_{Kinetic} = \bar{f}\gamma^{\mu}(iD_{\mu})f + (i\partial_{\mu}\phi_{2})^{\dagger}(i\partial^{\mu}\phi_{2}) + (i\partial_{\mu}\phi_{3})^{\dagger}(i\partial^{\mu}\phi_{3})
\label{K}
\ee
with $D_{\mu} = \partial_{\mu} -i\frac{Y}{2}g^{'}B_{\mu}$ and
\be
\begin{split}
V(H,\phi_{2},\phi_{3}) = m^{2}H^{\dagger}H + m_{2}^{2}\phi_{2}^{\dagger}\phi_{2} + m_{3}^{2}\phi_{3}^{\dagger}\phi_{3} + \lambda /2(H^{\dagger}H)^{2} + \lambda_{2} /2(\phi_{2}^{\dagger}\phi_{2})^{2} + \lambda_{3} /2(\phi_{3}^{\dagger}\phi_{3})^{2}\\
+ m_{13}(H^{\dagger}H)\phi_{3} + m_{23}(\phi^{\dagger}_{2}\phi_{2})\phi_{3} + \mu_{3}\phi^{3}\\
+ \lambda_{12}(H^{\dagger}H)(\phi^{\dagger}_{2}\phi_{2}) + \lambda_{13}(H^{\dagger}H)(\phi^{\dagger}_{3}\phi_{3}) + \lambda_{23}(\phi^{\dagger}_{2}\phi_{2})(\phi^{\dagger}_{3}\phi_{3}).
\end{split}
\label{scalar-pot}
\ee
Now since we require that the VEV of $\phi_{2}$ to be zero, that can be guaranteed if $m_{2} = 0$ and $m_{23}$, $\lambda_{12}$, $\lambda_{23}$ and $\lambda_{2}$ are all real and of same sign. If LHC is sensitive to the Higgs decays such as $h \rightarrow$ missing energy, then the couplings $\lambda_{12}$ and $\lambda_{13}$ can introduce terms such as $\lambda_{12}(v_{0} + h)^{2}\phi_{2}^{\dagger}\phi_{2}$ and $\lambda_{13}(v_{0} + h)^{2}(v_{3} + h_{3})^{2}$ which, beside contributing to the masses of $\phi_{2}$, $\phi_{3}$ and Higgs itself, can induce Higgs decay into missing energy i.e $h \rightarrow$ missing energy $(\bar{\phi_{2}}\phi_{2})$ and $h \rightarrow$ missing energy $(\bar{h_{3}}h_{3})$ provided $m_{h} > 2m_{\phi_{3}}$, where $v_{0}$ and $v_{3}$ are the VEV of the SM Higgs and $\phi_{3}$ respectively. One main constrain on the mass of the $h_{3}$ comes from the on shell Z decay $Z \rightarrow \gamma h_{3}$ via the triangle loop, which can affect the Z width, since no deviation has been reported in the Z decay width, we can avoid the decay if we require $m_{h_{3}} > m_{Z}$.\\
\\
With the $U(1)_{Y}$ gauge boson $B_{\mu}$ expressed in terms of $Z_{\mu}$ and $A_{\mu}$ as
\be
B_{\mu} = -\sin{\theta_{W}}Z_{\mu} + \cos{\theta_{W}}A_{\mu},
\ee
and from Eqs.(\ref{K}) we can express the interaction of $Z_{\mu}$ and $A_{\mu}$ with $\bar{f}\gamma^{\mu}f$ current as
\be
\bar{f}\gamma^{\mu}i(-i\frac{Y}{2}g^{'}B_{\mu})f = -e\bar{f}\gamma^{\mu}(-\tan{\theta_{W}}Z_{\mu} + A_{\mu})f,
\label{Eqs-A-Z}
\ee
where $g^{'} = \frac{e}{\cos{\theta_{W}}}$. From the above equation we can see that this heavier muon can be produced in colliders as $e^{+}e^{-} \rightarrow f^{+}f^{-}$ at $e^{+}e^{-}$ colliders and as $pp \rightarrow Z^{*}/\gamma^{*} \rightarrow f^{+}f^{-}$ at LHC.

\subsection{Scalar quartic couplings.}

We impose the conditions $m_{2} = 0$ and $m_{23}$, $\lambda_{12}$, $\lambda_{23}$ and $\lambda_{2}$ are all real and of same sign, to make sure the $\phi_{2}$ does not develop a non zero VEV. Since LHC is sensitive to Higgs decays such as $h \rightarrow missing\ energy\ (\bar{\phi_{2}}\phi_{2})$ and $h \rightarrow missing\ energy\ (\bar{h_{3}}h_{3})$, the couplings $\lambda_{12}$ and $\lambda_{13}$ can be probed at LHC provided $m_{h} > 2m_{\phi_{3}},\ 2m_{\phi_{2}}$. Given that LHC did not found any excess in the invisible Higgs decay over the SM background, the special case where the couplings $m_{13}$, $\lambda_{12}$ and $\lambda_{13}$ are also very small is favored by the present LHC data. So then neglecting terms containing $m_{13}$, $\lambda_{12}$ and $\lambda_{13}$ in Eqs.(\ref{scalar-pot}), we can write the scalar potential as
\be
\begin{split}
V(H,\phi_{2},\phi_{3}) = m^{2}H^{\dagger}H + m_{3}^{2}\phi_{3}^{\dagger}\phi_{3} + \lambda /2(H^{\dagger}H)^{2} + \lambda_{2} /2(\phi_{2}^{\dagger}\phi_{2})^{2} + \lambda_{3} /2(\phi_{3}^{\dagger}\phi_{3})^{2}\\
+ m_{23}(\phi^{\dagger}_{2}\phi_{2})\phi_{3} + \mu_{3}\phi^{3} + \lambda_{23}(\phi^{\dagger}_{2}\phi_{2})(\phi^{\dagger}_{3}\phi_{3}).
\end{split}
\ee
In this limit, the SM Higgs completely decouple from the other scalars and the scalar potential of $\phi_{3}$ is similar to SM Higgs potential except the term $\lambda_{23}(\phi^{\dagger}_{2}\phi_{2})(\phi^{\dagger}_{3}\phi_{3})$. This term give mass to $\phi_{2}$ after $\phi_{3}$ develop a non zero VEV $v_{3}$, where we have $\phi_{3} = v_{3} + h_{3}$ with $\phi_{3}$ being a real scalar.

\subsection{Contribution to $(g-2)_{\mu}$ in the model.}

In \cite{Leveille}, the contribution from a scalar ($\phi$) and a charged lepton f to the muon anomalous magnetic moment has been calculated in an arbitrary gauge for an interaction Lagrangian given as
\be
\mathcal{L} = \bar{\mu}(C_{S} + C_{P}\gamma^{5})f\phi,
\ee
and they give in Eqs.(11) of \cite{Leveille}
\be
[a_{\mu}]_{\phi} = \frac{-q_{f}m^{2}_{\mu}}{8\pi^{2}}\int^{1}_{0}dxQ_{\phi}(x),
\ee
where
\be
Q_{\phi}(x) = \frac{[ C_{S}^{2}\{ x^{2} - x^{3} + \frac{m_{f}}{m_{\mu}}x^{2} \} + C_{P}^{2}\{ m_{f} \rightarrow -m_{f} \} ]}{m_{\mu}^{2}x^{2} + (m_{f}^{2} - m_{\mu}^{2})x + m_{\phi_{2}}^{2}(1 - x)}.
\ee
For the case relevant to the model in this work the above formula reduces to
\be
[a_{\mu}]_{NP} = \frac{m^{2}_{\mu}Y_{2}^{2}}{16\pi^{2}}\int^{1}_{0}dx\frac{x^{2} - x^{3}}{m_{\mu}^{2}x^{2} + (m_{f}^{2} - m_{\mu}^{2})x + m_{\phi_{2}}^{2}(1 - x)},
\ee
where we have $C_{S} = -C_{P} = \frac{Y_{2}}{2}$ and $q_{f} = -1$. Now $m_{\phi_{2}} \geq m_{f}$ is not allowed because then we will have a stable long live charged particle ($f^{\pm}$) in our model, which will contradict the non-observations of any signatures from such a heavy long lived charged particles in the past experiments. So a reasonable approximation may seem to assume $m_{f} >> (m_{\mu}, m_{\phi_{2}})$, which allow $\phi_{2}$ to be a possible dark matter candidate. With this condition on the masses, the NP contribution to the muon anomalous magnetic moment is given as
\be
\delta a_{\mu} = [a_{\mu}]_{NP} \approx \frac{m^{2}_{\mu}}{16\pi^{2}}\frac{Y_{2}^{2}}{m_{f}^{2}}(\frac{1}{2} - \frac{1}{3}) = \frac{m^{2}_{\mu}}{6\times 16\pi^{2}}\frac{Y_{2}^{2}}{m_{f}^{2}}.
\ee
Then from the 1$\sigma$ of the central value of $\delta a_{\mu}$ from the Eqs.(\ref{diff-a}) we get
\be
\frac{Y_{2}}{m_{f}} = 0.0133\ GeV^{-1},
\ee
and from the perturbativity condition, i.e $Y_{2} \leq 1$, within 1$\sigma$ of the central value of $\delta a_{\mu}$, the maximum allowed value of $m_{f}$ is 75.19 GeV. Non observations of tracts of any long lived heavy charged particles in the past experiments indicates that $m_{f}$ should be close to this upper limit, as then its Yukawa coupling would be large, and it could have decayed immediately after its production and no observable tracts are left. However in this case since $m_{Z} > 2m_{\phi_{2}}$, the Z decay $Z \rightarrow \phi_{2}\phi_{2}$, via the triangle loop, can occur and expected to be large and so this case is already ruled out by the total Z decay width which is consistent with SM. For the case where $m_{f} \sim m_{\phi_{2}} >> m_{\mu}$, within 1$\sigma$ of the central value of $\delta a_{\mu}$, we get $\frac{Y_{2}}{m_{f}} = 0.0188 GeV^{-1}$ with maximum allowed value of $m_{f}$ (and also $\sim m_{\phi_{2}}$) from the perturbativity is 53.19 GeV, so $Z \rightarrow \phi_{2}\phi_{2}$ is forbidden by kinematics, and the condition $m_{f} > m_{\phi_{2}} + m_{\mu}$ imposed so that there is no heavy stable charged particle in the model in this mass range.

\subsection{Small neutrino masses and a dark-matter.}

If we add three right handed SM singlets $\nu_{iR}$ to the three left handed neutrinos in SM, then neutrinos can have Dirac mass term given as\\
\be
\mathcal{L}^{D}_{M} = \bar{\nu}_{R}M_{d}\nu_{L} + h.c
\ee
It has been borne out by many experimental observations and theoretical estimates that neutrinos have very small masses of $\mathcal{O}(10^{-10})$ Gev. The particle content of the model given in this work allows us to write more general Yukawa couplings for the neutrinos given as
\be
Y_{s_{ij}}\bar{\nu}_{R}^{i}s_{L}^{j}\phi_{3} + Y_{\eta_{ij}}\bar{s^{c}}_{iL}s_{jL}\phi_{3} + h.c
\label{majorana-mass}
\ee
where indice $i, j$ refers to the lepton generation and $s_{iL}$ are neutral leptons which are singlet under the SM gauge groups. When $\phi_{3}$ develops a non zero VEV, these Yukawa interaction terms can give masses to the respective fermions and can provide a simple explanation for the smallness of neutrino mass through the inverse-seesaw mechanism. In inverse-seesaw, on top of Dirac neutrino mass $m_{\nu_{l}}$, to generate small neutrino mass, we introduce a new left handed SM singlet neutral fermion $s_{iL}$ having a Dirac mass $M_{s}$ with $\nu_{R}^{l}$ and a small lepton number breaking Majorana mass $\eta$ (inverse seesaw mechanism) \cite{inverse-seesaw}, with the final mass term given as
\be
\mathcal{L}_{inverse-seesaw} = \bar{\nu_{R}}M_{d}\nu_{L} + \bar{\nu_{R}}M_{s}s_{L} + \bar{s^{c}_{L}}\eta s_{L} + h.c
\label{Invse-mass}
\ee
where $M_{\nu} = v_{0}Y_{Dirac}$, $M_{s} = v_{3}Y_{s}$ and $\eta = v_{3}Y_{\eta}$ are $3\times 3$ mass matrices with $v_{0}$ SM Higgs VEV and $v_{3}$ is the VEV of $\phi_{3}$. In the simplest case we can take the mass matrices such that $U_{R}^{\dagger}M_{d}U_{L} = diag(m_{d_{1}},m_{d_{2}},m_{d_{3}})$, $U_{R}^{\dagger}M_{s}O_{s} = diag(m_{s_{1}},m_{s_{2}},m_{s_{3}})$ and $O_{s}^{T}\eta O_{s} = diag(\eta_{1},\eta_{2},\eta_{3})$ where $U_{R}$ and $U_{L}$ are the usual unitary matrices that diagonalize the Dirac mass matrix while $O_{s}$ is an orthogonal matrix that diagonalize the $\eta$ mass matrix with $M_{s}$ such that it is diagonalized by a unitary matrix $U_{R}^{\dagger}$ from left and an orthogonal matrix $O_{s}$ from the right and the indice 1, 2 and 3 refers to the lepton generation. We would like to point out that although it is not necessary that acting matrix $U_{R}^{\dagger}$ from left and $O_{s}$ from right also diagonalize the matrix $M_{s}$, but if indeed it does diagonalize the matrix $M_{s}$ as assumed above, then the structure and mechanism of small neutrino masses of three generations split into three identical $2\times2$ matrix as shown in Eqs.(\ref{Eqs:Mass-Matrix}), i.e, in this model it is possible to make the neutrino mass generation mechanism universal among the three neutrino flavors. Then Eqs.(\ref{Invse-mass}) reduces to
\be
\mathcal{L}_{inverse-seesaw} = \sum_{i = 1}^{3}(\bar{\nu}_{Ri}\ \bar{s}^{c}_{Li})\left(\begin{array}{cc} m_{d_{i}} & m_{s_{i}} \\ 0 & \eta_{i} \end{array} \right) \left(\begin{array}{c} \nu_{Li} \\ s_{Li} \end{array} \right) + h.c,
\label{Eqs:Mass-Matrix}
\ee
and each of these mass matrices can be diagonalized as
\be
\left(\begin{array}{cc} m_{h_{i}} & 0 \\ 0 & m_{l_{i}} \end{array} \right) = \left(\begin{array}{cc} \cos(\lambda_{i}) & \sin(\lambda_{i}) \\ -\sin(\lambda_{i}) & \cos(\lambda_{i}) \end{array} \right) \left(\begin{array}{cc} m_{d_{i}} & m_{s_{i}} \\ 0 & \eta_{i} \end{array} \right) \left(\begin{array}{cc} \cos(\lambda_{i}) & -\sin(\lambda_{i}) \\ \sin(\lambda_{i}) & \cos(\lambda_{i}) \end{array} \right).
\ee
Then the smallness of the light neutrino mass for each generation comes from smallness of $\eta_{i}$ and is given as \cite{Park}
\be
m_{l_{i}} \approx \eta_{i}\frac{m_{d_{i}}^{2}}{m_{d_{i}}^{2} + m_{s_{i}}^{2}} \le \mathcal{O}(10^{-10})\ GeV\ and\ m_{h_{i}} \approx \frac{m_{d_{i}}^{2} + m_{s_{i}}^{2}}{m_{d_{i}}},
\ee
where $\eta_{i}$ can be taken at the scale of the breaking of $U(1)_{Lepton}$(Lepton number) with $\sin(\lambda_{i}) = \frac{m_{s_{i}}}{\sqrt{m_{d_{i}}^{2} + m_{s_{i}}^{2}}}$ and $\cos(\lambda_{i}) = \frac{m_{d_{i}}}{\sqrt{m_{d_{i}}^{2} + m_{s_{i}}^{2}}}$ and $m_{l_{i}}$ and $m_{h_{i}}$ refers to the masses of light and heavy neutrinos respectively. In the limit $m_{s_{i}} >> m_{d_{i}}$, we have $\nu_{l_{i}} = -\sin(\lambda_{i})(\nu_{Li} + \nu_{Ri}) + \cos(\lambda_{i})(s^{c}_{Li} + s_{Li}) \approx -\mathcal{O}(1)(\nu_{Li} + \nu_{Ri}) + \mathcal{O}(\frac{m_{d_{i}}}{m_{s_{i}}})(s^{c}_{Li} + s_{Li}) \approx -\nu_{i}$ and similarly $\nu_{h_{i}} \approx \mathcal{O}(1)(s_{Li} + s_{Li}^{c}) + \mathcal{O}(\frac{m_{d_{i}}}{m_{s_{i}}})$ where $\nu_{l_{i}}$ is the light neutrino and $\nu_{h_{i}}$ is the heavy neutrino for each generation denoted by the index $i$. In this limit the light neutrinos mainly consist of $\nu_{Li} + \nu_{Ri}$ and the heavy neutrinos consist mainly of $s^{c}_{Li} + s_{Ri}$. Then we have $\nu_{Li} \approx -P_{L}\nu_{li} + \mathcal{O}(\frac{m_{di}}{m_{si}})P_{L}\nu_{hi}$ and so corrections to neutrino oscillation due to heavy neutrino ($\nu_{hi}$) is at the order of $\mathcal{O}(\frac{m_{di}}{m_{si}})$ where $P_{L} = \frac{1}{2}(1 - \gamma^{5})$. Due to interactions in Eqs.(\ref{majorana-mass}) and Eqs.(\ref{Yukawa}), $h_{3}$ can decay into light neutrinos as $h_{3} \rightarrow \bar{\nu}\nu$ and photons as $h_{3} \rightarrow \gamma\gamma$ (via the triangle loop) respectively, and so only $\phi_{2}$ will be a stable neutral scalar that can contribute to the Dark-Matter (DM) of the universe. In the case where $m_{f} \sim m_{\phi_{2}} >> m_{\mu}$, within 1$\sigma$ of the central value of $\delta a_{\mu}$, the mass of the scalar DM ($m_{\phi_{2}}$) is allowed to be as high as about 53 GeV. But it turns out that for large Yukawa couplings, such as in the case of our model, the DM annihilation crossection is too large that $\phi_{2}$ could only contribute a very small fractions of the DM relic density of the universe, see \cite{Abe}\cite{Chaing} for detail calculations in these type of models.

\section{Production and signature in future searches.}

Since the $\bar{f}f$ current interact with electromagnetic and neutral weak gauge bosons $\gamma$ and $Z$ respectively as given in Eqs.(\ref{Eqs-A-Z}), we can produce $f^{+}f^{-}$ pair in the colliders as
\be
e^{+}e^{-} \rightarrow \gamma^{*}/Z^{*} \rightarrow f^{+}f^{-}
\ee
at $e^{+}e^{-}$ colliders and as
\be
pp \rightarrow \gamma^{*}/Z^{*} \rightarrow f^{+}f^{-}
\ee
at LHC. But due to non-observation of tracts of long lived charged particles in the collider experiments in the past, it is reasonable to assume that the the heavy lepton f are very short lived particle, i.e $Y_{2} \approx 1$, and so according to $\frac{Y_{2}}{m_{f}} = 0.0188\ GeV^{-1}$, its mass should be close to maximum allowed value of about 53.19 GeV. Then the heavy lepton decay very quickly after its production into $f^{\pm} \rightarrow \mu^{\pm}\phi_{2}$ and it would have left no tracts, but since $\phi_{2}$ is stable, light and very long lived neutral particle, possible candidate for DM, it would have been mistaken with the direct production of muon pair as $e^{+}e^{-}/pp \rightarrow \gamma^{*}/Z^{*} \rightarrow \mu^{+}\mu^{-}$ especially at LHC where only part of the energy carried by protons estimated from Parton-distribution-functions (PDF) is transferred to the final $\mu^{+}\mu^{-}$ state. However at $e^{+}e^{-}$ colliders, since the total energy carried by the $e^{+}e^{-}$ pair are transferred to $f^{+}f^{-}$ pair in full, we can look for key signal at $e^{+}e^{-}$ colliders as
\be
e^{+}e^{-} \rightarrow \gamma^{*}/Z^{*} \rightarrow f^{+}f^{-} \rightarrow \mu^{+}\mu^{-} + missing\ energy\ (\phi_{2} \phi_{2}).
\ee
In the $e^{+}e^{-}$ colliders, the final state $\mu^{+}\mu^{-} + missing\ energy\ (\phi_{2} \phi_{2})$ of NP can be differentiated from the SM final state $\mu^{+}\mu^{-}$ pretty easily by measuring the missing energy   in the final $\mu^{+}\mu^{-}$ state relative to the total energy of the initial $e^{+}e^{-}$ pair, and so $e^{+}e^{-}$ colliders running at 110 GeV or higher center of mass energies are very suitable to detect the presence of $f^{+}f^{-}$ particles.
Another key signature at LHC would be the triangle loop final state
\be
pp \rightarrow \gamma^{*}/Z^{*} \rightarrow \gamma + \phi_{3} \rightarrow \gamma + (\phi_{3} \rightarrow \bar{\nu}\nu\ or\ \bar{\phi_{2}}\phi_{2}) \rightarrow \gamma + missing\ energy.
\ee
In case of fermion f affecting the Z decay $Z \rightarrow \bar{\mu}\mu$ via the triangle loop \cite{Chaing}, we have for $m_{f} = 53.19$ GeV and $m_{\phi_{2}} = 53$ GeV, $Br(Z \rightarrow \bar{\mu}\mu)_{NP} = 2.19\times 10^{-6}$ compare to the PDG average \cite{pdg-Z} of $Br(Z \rightarrow \bar{\mu}\mu)_{Exp} = (3.366 \pm 0.007)\%$, so NP contribution is an order of magnitude smaller than the experimental error.\\
Similarly the other heavy leptons carrying electron lepton number ($f_{e}$) and tau lepton number ($f_{\tau}$) can exist and due to smallness of electron mass relative to moun mass it is easy to accommodate the fact that SM accounts very well for the $a_{e}$; for instance if we take $m_{f_{e}} \approx m_{f_{\mu}}$ and $Y_{e} = 1$, then $\delta a_{e}^{NP} \approx \mathcal{O}(5\times 10^{-14})$ compared to the error in the experimental estimation of $\delta a_{e}^{Exp-Error} \approx \mathcal{O}(2.6\times 10^{-13})$ \cite{pdg-muon}. For the $f_{\tau}$ the parameters are even less constrained because $a_{\tau}$ is not measured accurately yet.

\section{Conclusions.}

In this work we proposed a simple extension of SM by adding eleven additional new particles to it, three heavy lepton ($f_{e},\ f_{\mu},\ f_{\tau}$) only charged under the SM gauge group $U(1)_{Y}$ with $Y = -2$ and $f \rightarrow -f$ under a discrete symmetry transformation carrying respective Lepton Numbers. One scalar $\phi_{2}$, singlet under all the SM gauge groups and $\phi_{2} \rightarrow -\phi_{2}$ under the discrete symmetry with zero VEV. One more scalar $\phi_{3}$, singlet under all the SM gauge groups and invariant under the discrete symmetry which develops a non zero VEV ($v_{3}$) and participates in the mass generations of $\phi_{2}$, $f_{i}$s, neutrinos and SM Higgs as well as it self. Three right-handed neutrinos ($\nu_{iR}$) and three left-handed Majorana neutrinos ($s_{iL}$). With these eleven additional particles added to SM we have been able to give explanations to the long standing $(g -2)_{\mu}$ anomaly as well as the smallness of neutrino masses by the inverse see-saw mechanism. And also in this model we have a very suitable scalar dark-matter (DM) candidate in $\phi_{2}$ with allowed mass as high as about 53 GeV, although due to large Yukawa coupling from muon (g-2), its contribution to the DM relic density turn out to be too small and so it can account only a small fraction of the DM relic density of the universe. And we have also analyzed some possible means of production and signature that can show up in LHC and present and future $e^{+}e^{-}$ colliders. We proposed the main production and signature can be found in the processes $e^{+}e^{-}/pp \rightarrow f^{+}f^{-} \rightarrow \mu^{+}\mu^{-} + missing\ energy\ (\bar{\phi_{2}}\phi_{2})$ and $pp \rightarrow \gamma^{*}/Z^{*} \rightarrow \gamma + missing\ energy\ (\phi_{3} \rightarrow \bar{\nu}\nu\ or\ \bar{\phi_{2}}\phi_{2})$ via the triangle loop of fermions $f_{i}$. So discovery potential lies in the capability of a particular collider to measure the missing energy in the final state. In that sense we think $e^{+}e^{-}$ colliders are much better suited for discovery of the signatures of these new particles than LHC. In $e^{+}e^{-}$ colliders, initial total energy in the CM of the $e^{+}e^{-}$ pair is transferred to the final $\mu^{+}\mu^{-}$ pair in full, we can detect the missing energy easily where as at LHC, only partons in the protons participate whose energy are only known in terms of PDF and therefore its sensitivity towards missing energy will be reduced.

{\large Acknowledgments: \large} This work is supported and funded by the Department of Atomic Energy of the Government of India and by the Government of Tamil Nadu and Government of U.P.

\end{document}